\documentclass[10pt,english]{article}
\usepackage[T1]{fontenc}
\usepackage[latin9]{inputenc}
\usepackage[a4paper]{geometry}
\usepackage{array}
\usepackage{float}
\usepackage{amstext}

\makeatletter

\providecommand{\tabularnewline}{\\}

\usepackage[left,modulo]{lineno}
\usepackage{multicol} 
\usepackage{lmodern}

\makeatother

\usepackage{babel}
\begin{document}

\title{{\huge{An Analysis of issues against the adoption of Dynamic Carpooling}}}

\author{{\Large{Daniel Graziotin, Free University of Bozen-Bolzano, daniel.graziotin@unibz.it}}}

\date{{\Large{December 2010}}}
\maketitle
\begin{abstract}
Using a private car is a transportation system very common in industrialized
countries. However, it causes different problems such as overuse of
oil, traffic jams causing earth pollution, health problems and an
inefficient use of personal time. 

One possible solution to these problems is carpooling, i.e. sharing
a trip on a private car of a driver with one or more passengers. Carpooling
would reduce the number of cars on streets hence providing worldwide
environmental, economical and social benefits. The matching of drivers
and passengers can be facilitated by information and communication
technologies. Typically, a driver inserts on a web-site the availability
of empty seats on his/her car for a planned trip and potential passengers
can search for trips and contact the drivers. This process is slow
and can be appropriate for long trips planned days in advance. We
call this static carpooling and we note it is not used frequently
by people even if there are already many web-sites offering this service
and in fact the only real open challenge is widespread adoption.

Dynamic carpooling, on the other hand, takes advantage of the recent
and increasing adoption of Internet-connected geo-aware mobile devices
for enabling impromptu trip opportunities. Passengers request trips
directly on the street and can find a suitable ride in just few minutes.
Currently there are no dynamic carpooling systems widely used. Every
attempt to create and organize such systems failed.

This paper reviews the state of the art of dynamic carpooling. It
identifies the most important issues against the adoption of dynamic
carpooling systems and the proposed solutions for such issues. It
proposes a first input on solving the problem of mass-adopting dynamic
carpooling systems.
\end{abstract}

\section{Introduction}

Using a private car is a transportation system very common in industrialized
countries. Between 2004 and 2009, the worldwide production of private
vehicles has been of 295 millions of new units %
\footnote{(Accessed Dec 13 2010) http://oica.net/%
} and, as of 2004, there were 199 millions registered drivers in the
U.S.A.%
\footnote{ (Accessed Dec. 13 2010) http://www.fhwa.dot.gov/ {[}U.S. Department
of Transportation - Federal Highway Administration{]}%
}. Road transport is responsible for about 16\% of man-made CO2 emissions%
\footnote{(Accessed Dec. 13 2010) http://oica.net/ {[}Organisation Internationale
des Constructeurs d\textquoteright{}Automobile{]}%
}. 

Private car travelling is a common but wasteful transportation system.
Most cars are occupied by just one or two people. Average car occupancy
in the U.K. is reported to be 1.59 persons/car, in Germany only 1.05
\cite{key-0}. Private car travelling creates a number of different
problems and societal costs worldwide. Environmentally, it is responsible
for a wasteful use of a scarce and finite resource, i.e. oil, and
causes unnecessary earth pollution. The traffic caused by single occupancy
vehicles causes traffic jams and hugely increases the amount of time
spent by people in queues on streets. This is a unsavvy use of another
scarce resource: time. Moreover, the additional pollution creates
health problems to millions of individuals. Lastly, lone drivers in
separate cars miss opportunities to meet and talk, incurring in a
loss of potential social capital.

One possible solution to all these problems is carpooling, i.e. the
act of sharing a trip on a private vehicle between one or more other
passengers. The shared use of a single car by two or more people would
reduce the number of cars on streets. Carpooling helps the environment
by allowing to use oil wisely, to reduce earth pollution and consequent
health problems. It reduces traffic and - consequently - time that
people spend in their cars. Carpooling has also the potential of increasing
social capital by letting people meet and know each other.

Carpooling is not a widespread practice. There are already many systems
facilitating the match between drivers and passengers, most of them
in form of bulletin board-like web-sites. The intention of offering
empty seats of a vehicle is usually announced by a driver many days
before the start of the trip. The coordination between a driver and
the passengers who are candidating for sharing the trip with him/her
is usually carried out by e-mails or private messages in the web-site.
Therefore, we may see carpooling as a \emph{static} way of sharing
a trip.

The availability of geo-aware, mobile devices connected to the Internet
opens up possibilities for the formation of carpools in short notice,
directly on streets. This phenomenon is called dyamic carpooling (also
known as dynamic ridesharing, instant ridesharing and agile ridesharing).
Dan Kirshner, researcher in this field and maintainer of http://dynamicridesharing.org
website defines it as follows: ``A system that facilitates the ability
of drivers and passengers to make one-time ride matches close to their
departure time, with sufficient convenience and flexibility to be
used on a daily basis.''%
\footnote{Kirshner, D. (Accessed Dec 10$^{\text{th}}$ 2010) - http://dynamicridesharing.org %
}.

Currently there are no dynamic carpooling systems widely used. In
fact, there are many problematic issues related to the implementation
and the adoption of dynamic carpooling systems. In Section 2 we present
our analysis of the state of the art on dynamic carpooling. We collected
all research papers about dynamic carpooling and issues against its
mass adoption. We critically analyze the issues and the solution proposals
in Section 3 of this paper. While we acknowledge all aspects are critical,
we claim that the basic technological infrastructure is an important
required and key building block. In Section 4 we summarize our thoughts
and outcomes about this analysis.

\subsection{Terminology}

In Table 1 we introduce some key concepts used in this paper.

\begin{table}[H]
\begin{centering}
\begin{tabular}{|c|>{\raggedright}p{0.6\textwidth}|}
\hline 
Term & Definition\tabularnewline
\hline 
\hline 
Person & \multicolumn{1}{l|}{A user registered in the system, with login and password}\tabularnewline
\hline 
Trip & The Driver can create Trips in the system. A Trip is the information
about the availability of seats in a car going from a certain location
to a certain destination, driven by a Driver on a certain date\tabularnewline
\hline 
Driver & The role of a Person when he/she offers to share some seats on his/her
car for a specific trip\tabularnewline
\hline 
Passenger & The role of a Person when he/she accepts to occupy a seat on a car
of a Driver\tabularnewline
\hline 
Participation & The act of taking part into a Trip. The Driver participates by default
in a Trip he/she created. A Passenger can participate in Trips created
by a Driver. Participation can be just requested by the passenger
or already confirmed by the driver.\tabularnewline
\hline 
Location & A geographical location.\tabularnewline
\hline 
\end{tabular}
\par\end{centering}

\caption{Key Terms of this paper}
\end{table}

\section{State of the art}

This section contains a summary of previously published papers, in
order of publication. In the next section we present the outcomes
of the analysis of the whole state of the art and how we decided to
move in order to provide a significant contribute in solving the problem
of adopting dynamic ridesharing services.

\subsubsection*{Sociotechnical support for Ride Sharing\cite{key-3}}

This paper lists barriers to dynamic carpooling adoption and possible
actions to reduce them. It reports about High Occupancy Vehicles (HOV)
lane - which are lanes dedicated for people doing carpooling - on
streets of San Francisco and Oakland and complains that there should
be no fees on bridges for HOVs. The author suggests conventions developed
between drivers and passengers (e.g. pickup points near public transportation
stops). Regarding security, the paper suggests to give priority to
female passengers, to not leave them alone waiting for a ride. The
paper reports that there are no stories about rape, kidnapping or
murder and the most common reported problem is bad driving.

There are suggestions on needed research: 
\begin{itemize}
\item Need of location-aware devices, because dynamic carpooling is actually
limited to fixed pickups and drop-off locations. 
\item Simple user interfaces for passengers and drivers. 
\item Routing matching algorithms: short window of opportunity to match
passenger and driver. 
\item Time-to-pickup algorithms: to help passenger decide whether to use
carpooling or Public Transportation System. 
\item Safety and reputation system design: authenticate passenger and driver
before making the match, monitor arrival at destination, feedback
system. 
\end{itemize}
The paper discusses about social capital impacts: there is the potential
for creating new social connections and also matching drivers and
passengers according to their profiles creates bridging across class,
race and religious views.

\subsubsection*{Pilot Tests of Dynamic Ridesharing\cite{key-4}}

The author presents three pilot tests done in the USA, all of them
failed. The reasons of failure are the following: 
\begin{itemize}
\item Too complicated rules and user interface
\item Too weak marketing effort
\item Too few users. After 1 month, 1000 flyers distributed to the public
and a proposed discount on parking, only 12 users were using the system. 
\end{itemize}
The paper adds the idea of saving money when parking. It also enforces
the idea of using social networks to allow car pooling on the fly.
The author envisions using a web \textendash{} and mobile service,
also introducing some interesting user stories.

\subsubsection*{The smart Jitney: Rapid, Realistic Transport \cite{key-5}}

The paper focuses on environmental benefits of dynamic carpooling.
It asserts that dynamic carpooling would lower greenhouse gas emissions
in a better way than electric/hydrogen/hybrid cars would do. It introduces
the idea of Smart Jitney: an unlicensed car driving on a defined route
according to a schedule.

The author suggests the installation of Auto Event Recorders on cars,
enforcing security. It complains that challenges are all focused in
convincing the population to use the service, proposing a cooperative
public development of the system.

\subsubsection*{Auction negotiation for mobile Rideshare service\cite{key-6}}

The paper proposes the use of agent-based systems powered auction
mechanisms for driver-passenger matching.

\subsubsection*{Casual Carpooling - enhanced\cite{key-10}}

The author considers areas without HOV lanes and proposes the use
of Radio Frequency IDentification (RFID) chips to quickly identify
passengers and drivers. Readers should be installed at common pick-up
points. The paper complains that it would cost less to pay passengers
and drivers for using the service than to build a HOV lane.

\subsubsection*{Empty seats travelling\cite{key-0}}

This white paper by Nokia suggests to use the phone as a mean of transportation,
creating a value in terms of a transport opportunity. It points out
some factors limiting static carpooling, arranged via websites: 
\begin{itemize}
\item Trip arrangements are not ad hoc
\item It is impossible to arrange trips to head home from work or to drive
shopping. 
\end{itemize}
The paper notices that people are not widely encouraged to practice
carpooling by local governments. It collects obstacles and success
factors in terms of human sentences, and their solution. The authors
say that the challenge is in the definition of a path leading from
existing ride share services to a fully automated system.

\subsubsection*{Interactive systems for real time dynamic multi hop carpooling\cite{key-11}}

The author proposes a dynamic multi-hop system, by dividing a passenger
route into smaller segments being part of other trips. The author
claims that the problems of static carpooling are that matching drivers
and passengers based on their destinations limits the number of possible
rides, and with high waiting times. Carpooling is static and does
not adapt itself well to ad hoc traveling. The paper asks governments
to integrate carpooling in laws and to push for its use. The author
complains that the perceived quality of service is increased even
driving the passenger away from destination: a driver and a passenger
should not be matched only if they share the same or similar destination
because perfect matching would require high waiting times. 

The paper also addresses social aspects: in a single trip with 3 hops
a passenger might meet 3 to 10 people, therefore passengers may be
socially matched. It suggests to link the application with some social
networks like Facebook, MySpace and use profile information to match
drivers and passengers. 

As security improvement, the paper suggests: the use of finger-prints,
RFID, voice signature, display the location of vehicles on a map,
using user pictures, assigning random numbers to be used as passwords.

\subsubsection*{Instant Social Ride Sharing\cite{key-12}}

The paper proposes matching methodologies based on both a minimization
of detours and the maximization of social connections. It assumes
the existence of a social network data source in which users are connected
by means of groups, interests, etc. In such a network, the number
of relatively short paths between a driver and a passenger indicates
the strength of their social connection.

It provides algorithms and SQL queries. The authors assume that there
is already a large scale of users, and no barriers to adoption are
taken into account.

\subsubsection*{Combining Ridesharing \& Social Networks\cite{key-13}}

The author envisions a mobile and web system that interacts with social
networks profiles that should improve security and trust by users.
Users can register to the system in a traditional way (e.g., by giving
email, username, password), then complete their profiles by linking
their accounts to multiple existing social networks account, to fill
the remaining fields. Otherwise, they have to fill the fields manually
and verify their identity in more classical ways. The paper proposes
Opensocial%
\footnote{Google, MySpace et al. (Accessed Dec. 19$^{\text{th}}$ 2010) - http://www.opensocial.org/%
} as connection interface. An own rating system is also complained,
which keeps scores of persons. Amongst the criteria are factors like
reliability, safety and friendliness. 

It suggests the use of mobile systems, that should make use of GPS
and creation of a match on the fly (real-time algorithms). The paper
provides some results of surveys: people are willing to loose 23\%
more time to pickup a friend of their social network rather than a
stranger (6\%). It also provides a high-level description of the system
and implementation details. 

The author asks for extra research on psychological factors that increase
trust and perceived safety.

\subsubsection*{SafeRide: Reducing Single Occupancy Vehicles \cite{key-14}}

The publication is about a project in the U.S.A. It reports that there
is a market-formation problem: to achieve the system that attracts
passengers, there will have to be many drivers available. But the
drivers will emerge only when it appears profitable or otherwise desiderable,
and that depends on there being many passengers, etc. The author complains
that someone must discover a winning formula before anyone will invest.

The paper lists some interesting user stories, as well as algorithms
and requirements.

\section{Comparative Analysis of Dynamic Carpooling Issues}

The analysis of the state of the art brought some issues related to
adopting dynamic carpooling systems. We categorized the issues gathered
from the state of the art and their proposals in the following categories:
\begin{itemize}
\item Interface Design - all issues related to graphical implementation
of clients and ease of use
\item Algorithms - the instructions regarding driver/passengers matching
problems
\item Coordination - the aspects related on how to let people meet, authenticate
and coordinate.
\item Trustiness - the problems related on raising user confidence on dynamic
carpooling systems 
\item Safety - the issues regarding ensuring protection of users
\item Social Aspects - all the issues related to create social connections
and raising social capital in dynamic carpooling systems 
\item Reaching Critical Mass - the problems on reaching a sufficient amount
of persons using the system that would attract more other people
\item Incentives - all the political, motivational and economical issues
related to dynamic ridesharing systems
\item System Suggestions - everything else that we consider relevant for
building dynamic carpooling systems
\end{itemize}
We attach the comparative analysis summarized in tables in Appendix,
Table 2 up to table 5. For each category (columns), we list the suggestions
and interesting points made in the different research papers (rows),
in form of imperative sentences. Tables 2 to 5 in Appendix is our
contribution rationalizing the many problematic issues involved in
the creation and deployment of dynamic carpooling systems and in summarizing
best practices and suggestions in how to deal with them. 

Our rationalization of dynamic carpooling issues and possible solutions
shows how dynamic carpooling systems still have many important open
issues to be addressed and solved. This fact explains the current
absence of any dynamic carpooling system deployed and used for real.
The most addressed issue is ``Reaching the critical mass''. This
problem has been faced in several ways but noone of them worked. This
issue seems very dependant to the issues named ``Incentives'', ``Safety''
and ``Trustiness''. In order to receive incentives from the government,
there must be a system providing safety. Incentives and safety would
produce positive feedbacks and provide trustiness among the general
public, therefore providing at least a palliative for the critical
mass issue. 

We decided to address the overall challenge from a very core point
of view and to focus on technical aspects. Among the projects, we
observed that their source code and the prototypes produced were not
freely accessible by the general public. There are no information
regarding the servers, that are all proprietary and obscured. Another
issue seems related to a missing standardization of the protocols
used. Therefore, every project started from zero, ``reinventing the
wheel''. In order to overcome the ``reaching critical mass'' issue,
we believe that it is important that providers of dynamic carpooling
services can exchange their data easily so that cross provider matching
are possible.

That is, the open problem of dynamic carpooling still needs the basic
building blocks of research on which future work should be performed
in order to overcome all the other issues.

\section{Conclusions and future work}

In this paper we presented the outcomes of a research aimed at providing
a better understanding to the open problem of dynamic carpooling.
Through an analysis of the state of the art (reported in Section 2),
we identified the key issues in the domain of dynamic carpooling,
presented in Section 3. Based on the comparative analysis of the open
issues, we notice that the basic building block is missing: an open
and extendable technological infrastructure. A future field of research
will be to create an opensource framework providing basic dynamic
carpooling functionalities and an open, discussable protocol. This
framework should be as much clear and extendable as possible, to let
other researchers work on different technical aspects. The other,
non technical issues should be solved afterwards.

Concluding, we believe the topic of this paper is a recent and challenging
one, still waiting for at least initial solutions and steps forwards.
Common goal of solving an important problem for our world: too many
cars on our streets with just one passenger in them.

\pagebreak{}

\section*{Appendix}

\begin{table}[H]
\begin{centering}
\begin{tabular}{|>{\raggedright}p{3cm}|>{\raggedright}p{4cm}|>{\raggedright}p{4cm}|>{\raggedright}p{4cm}|}
\hline 
Paper & Interface Design & Algorithms & Coordination\tabularnewline
\hline 
\hline 
\textit{Sociotechnical support for Ride Sharing}\textit{\footnotesize{\cite{key-3}}} & {\footnotesize{Give start, ending points and clear indications.}}{\footnotesize \par}

{\footnotesize{Filter what information to reveal}} &  & \tabularnewline
\hline 
\textit{Pilot Tests of Dynamic Ridesharing}\textit{\footnotesize{\cite{key-4}}} & {\footnotesize{Provide lots of flexible settings to satisfy users.}} &  & {\footnotesize{Provide a static/dynamic approach, let users insert
entries days before the start}}\tabularnewline
\hline 
\textit{The smart Jitney: Rapid, Realistic Transport }\textit{\footnotesize{\cite{key-5}}} & {\footnotesize{Provide different levels of services: }}{\footnotesize \par}

{\footnotesize{- simple: just destination and pickup }}{\footnotesize \par}

{\footnotesize{- groups preferences (only women etc.) }}{\footnotesize \par}

{\footnotesize{- scheduling of rides}} &  & \tabularnewline
\hline 
\textit{Auction negotiation for mobile Rideshare service}\textit{\footnotesize{\cite{key-6}}} &  &  & \tabularnewline
\hline 
\textit{Casual Carpooling - enhanced}\textit{\footnotesize{\cite{key-10}}} &  &  & {\footnotesize{Implement one-time registration process, simple.}}{\footnotesize \par}

{\footnotesize{Provide RFID devices for drivers and passengers}}\tabularnewline
\hline 
\textit{Empty seats travelling}\textit{\footnotesize{\cite{key-0}}} &  &  & \tabularnewline
\hline 
\textit{Interactive systems for real time dynamic multi hop carpooling}\textit{\footnotesize{\cite{key-11}}} & {\footnotesize{Focus on simplicity. Provide voice, speech recognition.
Allow users to communicate each other.}} &  & {\footnotesize{Driving passenger away from the destination but near
transportation locations (e.g. a bus station) increases quality of
service and enhances coordination.}}\tabularnewline
\hline 
\textit{Instant Social Ride Sharing}\textit{\footnotesize{\cite{key-12}}} &  & {\footnotesize{Given, built around social connections. Social network
needed.}} & {\footnotesize{Built around social connection between users}}\tabularnewline
\hline 
\textit{Combining Ridesharing \& Social Networks}\textit{\footnotesize{\cite{key-13}}} & {\footnotesize{Implement a simple registration system from mobile
phone. }}{\footnotesize \par}

{\footnotesize{In a second phase link social networks profiles, or
manual fill.}}{\footnotesize \par}

{\footnotesize{Develop a very simple UI}} &  & \tabularnewline
\hline 
\textit{SafeRide: Reducing Single Occupancy Vehicles}\textit{\footnotesize{\cite{key-14}}} &  & {\footnotesize{Both data structures and Algorithms for matching are
given}} & \tabularnewline
\hline 
\end{tabular}
\par\end{centering}

\caption{Paper Analysis: Interface Design, Algorithms, Coordination}
\end{table}

\pagebreak{}
\begin{table}[H]
\begin{centering}
\begin{tabular}{|>{\raggedright}p{3cm}|>{\raggedright}p{3cm}|>{\raggedright}p{3cm}|>{\raggedright}p{3cm}|}
\hline 
Paper & Trustiness & Safety & Social Aspects\tabularnewline
\hline 
\hline 
\textit{Sociotechnical support for Ride Sharing}\textit{\footnotesize{\cite{key-3}}} &  & {\footnotesize{Authenticate before the match: password / PIN monitor
arrival at destination}}{\footnotesize \par}

{\footnotesize{Provide a feedback system a la EBay}} & {\footnotesize{Announce matching items in profiles before the ride}}{\footnotesize \par}

{\footnotesize{Do research in social capital aspects}}\tabularnewline
\hline 
\textit{Pilot Tests of Dynamic Ridesharing}\textit{\footnotesize{\cite{key-4}}} &  & {\footnotesize{Create a PIN at registration phase to be used by the
client}} & {\footnotesize{Add social networking support to help finding neighbours}}\tabularnewline
\hline 
\textit{The smart Jitney: Rapid, Realistic Transport }\textit{\footnotesize{\cite{key-5}}} & {\footnotesize{Brand the idea: apply stickers on every car that participates. }}{\footnotesize \par}

{\footnotesize{Give limitations to drivers: age limits, extra driving
tests, check on criminal records etc.}} & {\footnotesize{Provide Auto Event Recorders on cars. Implement an
emergency button on mobile phone, record GPS data.}}{\footnotesize \par}

{\footnotesize{Provide a feedback system a la EBay}} & \tabularnewline
\hline 
\textit{Auction negotiation for mobile Rideshare service}\textit{\footnotesize{\cite{key-6}}} &  &  & \tabularnewline
\hline 
\textit{Casual Carpooling - enhanced}\textit{\footnotesize{\cite{key-10}}} & {\footnotesize{Record carpooling activity when cars pass through RFID
readers }} & {\footnotesize{Build it around RFID, record lots of data and positions}} & \tabularnewline
\hline 
\textit{Empty seats travelling}\textit{\footnotesize{\cite{key-0}}} & {\footnotesize{Involve community and governments in planning and implementation
phases}} & {\footnotesize{Let the service be available only to registered users;
Provide a Feedback system}} & {\footnotesize{Give the possibility to create social connections}}\tabularnewline
\hline 
\textit{Interactive systems for real time dynamic multi hop carpooling}\textit{\footnotesize{\cite{key-11}}} &  & {\footnotesize{Use RFID, GPS. Implement a complete rating system. }}{\footnotesize \par}

{\footnotesize{Display vehicle and driver information before entering
a vehicle. Display participants pictures. Assign random numbers for
passenger pickups to confirm the ride. }}{\footnotesize \par}

{\footnotesize{Provide voice and video features.}} & {\footnotesize{Match passengers socially.}}{\footnotesize \par}

{\footnotesize{Link the application to social networks.}}\tabularnewline
\hline 
\textit{Instant Social Ride Sharing}\textit{\footnotesize{\cite{key-12}}} & {\footnotesize{Use social networks to enhance it.}} &  & \tabularnewline
\hline 
\textit{Combining Ridesharing \& Social Networks}\textit{\footnotesize{\cite{key-13}}} & {\footnotesize{People are ready to spend 17\% more time to pickup
a friend of the social network rather than a stranger. Implement it.}} & {\footnotesize{Implement a rating system. Use and record GPS data.}}{\footnotesize \par}

{\footnotesize{Do extra research in this field.}} & \tabularnewline
\hline 
\textit{SafeRide: Reducing Single Occupancy Vehicles}\textit{\footnotesize{\cite{key-14}}} & {\footnotesize{Use social networks to enhance it.}} & {\footnotesize{Implement a GPS Help button. Record time, place, and
sound. Develop a Feedback system}} & \tabularnewline
\hline 
\end{tabular}
\par\end{centering}

\caption{Paper Analysis: Trustiness, Safety, Social Aspects}
\end{table}
\pagebreak{}
\begin{table}[H]
\begin{centering}
\begin{tabular}{|>{\raggedright}p{3cm}|>{\raggedright}p{3cm}|>{\raggedright}p{3cm}|>{\raggedright}p{3cm}|}
\hline 
Paper & Critical Mass & Incentives & Suggestions\tabularnewline
\hline 
\hline 
\textit{Sociotechnical support for Ride Sharing}\textit{\footnotesize{\cite{key-3}}} &  &  & {\footnotesize{Provide a location-aware system}}{\footnotesize \par}

{\footnotesize{Make use of mobile phones}}\tabularnewline
\hline 
\textit{Pilot Tests of Dynamic Ridesharing}\textit{\footnotesize{\cite{key-4}}} & {\footnotesize{Provide mass marketing before, during and after deployment.
Search for start-up incentives}} & {\footnotesize{Search an institutional sponsor.}}{\footnotesize \par}

{\footnotesize{Make the government provide parking spaces to participants}} & {\footnotesize{Implement both Web and mobile clients.}}{\footnotesize \par}

{\footnotesize{Implement a static and a dynamic approach. }}{\footnotesize \par}

{\footnotesize{Start with a many-to-one system: all at a single destination}}\tabularnewline
\hline 
\textit{The smart Jitney: Rapid, Realistic Transport }\textit{\footnotesize{\cite{key-5}}} &  & {\footnotesize{Use a cooperative, public development of the system}} & {\footnotesize{Implement a Web interface and mobile clients (using
phone calls)}}\tabularnewline
\hline 
\textit{Auction negotiation for mobile Rideshare service}\textit{\footnotesize{\cite{key-6}}} &  &  & \tabularnewline
\hline 
\textit{Casual Carpooling - enhanced}\textit{\footnotesize{\cite{key-10}}} &  & {\footnotesize{Make employers incentive employees. Involve Regional
Transportation Boards}} & \tabularnewline
\hline 
\textit{Empty seats travelling}\textit{\footnotesize{\cite{key-0}}} & {\footnotesize{Create an incremental service, starting from a thread
of backwards compatible services (bus, taxi). Don't introduce new
devices for the service, use mobile phones}} & {\footnotesize{Find a way to make the service a business case. Search
for public incentives}} & {\footnotesize{Implement the system mobile only. Record GPS data.}}{\footnotesize \par}

{\footnotesize{Provide a non-obtrusive system for authentication Research
on quality of service measures}}\tabularnewline
\hline 
\textit{Interactive systems for real time dynamic multi hop carpooling}\textit{\footnotesize{\cite{key-11}}} & {\footnotesize{A multi-hop system will solve the problem, as more
rides will be available, waiting times will decrease and quality will
rise.}} & {\footnotesize{Convince governments to change laws to enforce carpooling}} & {\footnotesize{Use a dynamic, multi-hop, real- time mobile system
to minimize waiting times, one hop at a time}}\tabularnewline
\hline 
\textit{Instant Social Ride Sharing}\textit{\footnotesize{\cite{key-12}}} &  &  & {\footnotesize{Use mobile phones and sms. Use GPS.}}{\footnotesize \par}

{\footnotesize{Use a provided high-level description of the system}}\tabularnewline
\hline 
\end{tabular}
\par\end{centering}

\caption{Paper Analysis: Critical Mass, Incentives, Suggestions Pt. 1}
\end{table}
\begin{table}[H]
\begin{centering}
\begin{tabular}{|>{\raggedright}p{3cm}|>{\raggedright}p{3cm}|>{\raggedright}p{3cm}|>{\raggedright}p{3cm}|}
\hline 
Paper & Critical Mass & Incentives & Suggestions\tabularnewline
\hline 
\hline 
\textit{Combining Ridesharing \& Social Networks}\textit{\footnotesize{\cite{key-13}}} & {\footnotesize{Involve users in some parts of development process.
Research further on this topic.}} &  & {\footnotesize{Implement a mobile and a web system that interacts
with social networks profiles.}}{\footnotesize \par}

{\footnotesize{Use Opensocial and other social networks.}}{\footnotesize \par}

{\footnotesize{Use our high level description of the whole system}}\tabularnewline
\hline 
\textit{SafeRide: Reducing Single Occupancy Vehicles}\textit{\footnotesize{\cite{key-14}}} & {\footnotesize{Market-formation problem: discover a new, winning formula.}}{\footnotesize \par}

{\footnotesize{Start with an existing service, like taxis.}}{\footnotesize \par}

{\footnotesize{Find large employers.}}{\footnotesize \par}

{\footnotesize{Serve events (i.e.. concerts)}} & {\footnotesize{Find money. Search for incentives from governments}} & {\footnotesize{Implement our Use Cases}}{\footnotesize \par}

{\footnotesize{Provide our functional requirements. Provide our non-functional
requirements}}\tabularnewline
\hline 
\end{tabular}
\par\end{centering}

\caption{Paper Analysis: Critical Mass, Incentives, Suggestions Pt.2}
\end{table}

\pagebreak{}

\end{document}